# Numerical simulation and experimental study of PbWO$_4$/EPDM and Bi$_2$WO$_6$/EPDM for the shielding of γ-rays

Chi Song [1,2] Jian Zheng [1*]  Quan-Ping Zhang[1]   Yin-Tao Li[1]   Ying-Jun Li[1]   Yuan-lin Zhou[1*]

[1] State Key Laboratory Cultivation Base for Nonmetal Composites and Functional Materials, Southwest University of Science and Technology, Mianyang 621010, China

[2] School of Materials Science and Engineering, Southwest University of Science and Technology, Mianyang 621010, China

**Abstract:** The MCNP5 code was employed to simulate the γ-ray shielding capacity of tungstate composites. The experimental results were applied to verify the applicability of the Monte Carlo program. PbWO$_4$ and Bi$_2$WO$_6$ were prepared and added into ethylene propylene diene monomer (EPDM) to obtain the composites, which were tested in the γ-ray shielding. Both the theoretical simulation and experiments were carefully chosen and well designed. The results of the two methods were found to be highly consistent. In addition, the conditions during the numerical simulation were optimized and double-layer γ-ray shielding systems were studied. It was found that the γ-ray shielding performance can be influenced not only by the material thickness ratio but also by the arrangement of the composites.

**PACS:** 23, 33

## 1 Introduction

Currently, radioactive nuclear physics is one of the most vibrant frontier research fields [1, 2]. With the application of nuclear science and technology in various fields, the requirements on nuclear safety have been progressively increased. Because of their great penetrative power, γ-rays can greatly harm the human body. They are electromagnetic waves with very short wavelength, which can be formed by heavy nuclear fission, fission product decay, radiative capture, inelastic scattering and activation product decay, etc [3]. Being exposed to them, the human body can be substantially damaged, with even cell carcinogenesis or genetic mutation. One of the key issues in the research of γ-ray shielding is to accurately understand the related parameters, such as characteristics of the radiation source, radiation energy, as well as the properties of the shielding materials.

Concrete, cement, lead boards and steel boards are utilized as conventional γ-ray shielding materials [4, 5]. However, they barely meet the demand in the industry of protective equipment due to the rapid growth of nuclear technology in more diverse areas. Since the 1970s, various special composite shielding materials have been developed. Those materials are usually made of organic polymer materials as base-materials and filled with a ray-absorptive substance [6]. In recent years, heavy metal particles, such as lead based materials, have also been filled into organic polymer base-materials as functional components to synthesize γ-ray shielding composites [7, 8]. Compared to Pb and PbO, which are highly toxic, lead and bismuth tungstate materials have a range of advantages, such as excellent stability, high performance in shielding, and non-toxic and harmless qualities [9].

In recent years, in collaborations with other groups, several tungstates for γ-ray shielding have been developed in our group [3, 9-14]. For instance, superfine lead tungstate powders were prepared, which were used as the functional additive and added into latex products to form the composites. The prepared composites exhibited excellent capacity in the attenuation of γ-rays [10, 11]. A hydrophobic lead tungstate particle whose average crystal size is 32 nm has been composited and used as functional filler to produce a γ-ray shielding paint-coat. It has been proved that the paint-coat has a shielding rate up to 32% [12]. The influence of particle size and crystalline morphology was also found to be an important parameter for the γ-ray shielding. PbWO$_4$ particles with different crystal sizes and morphologies were fabricated and employed as the functional additive to obtain the final composites. The results showed that the smaller the size of the $PbWO_4$ crystal, the better the dispersion in the latex, which is helpful to improve the shielding of the γ-rays [14].

Compared to experimental measurement, the theoretical simulation of γ-ray shielding possesses several benefits, such as high efficiency, lower cost and less damage to humans, and can be a powerful tool to support experiments [15]. For example, Sharifi et al. studied the shielding properties of concretes using MCNP-4C code and compared with available experimental results. The simulated and measured values were compared and the results showed reasonable agreement for all concretes [16]. Jalali et al. compared the experiment and computation of compounds NaBO, HBO, CdCl and NaCl and their solutions attenuate γ-rays in addition to neutron absorption. The results suggested that simulation by MCNP code has potential application for determining the attenuation coefficients of various compound materials [17].

$PbWO_4$ and $Bi_2WO_6$ are the two of the functional fillers which show the most potential for γ-ray shielding. However, there are only a few works focused on simulation of $PbWO_4$ and $Bi_2WO_6$ in γ-ray shielding, and comparing the difference between simulation and experiment. In this paper, MCNP code was used to calculate the $PbWO_4$ and $Bi_2WO_6$ potentially applied in γ-ray shielding. In addition, $PbWO_4$ and $Bi_2WO_6$ powder were also prepared by precipitation process and hydrothermal reaction and homogeneously dispersed in ethylene propylene diene monomer (EPDM), which was chosen as the substrate material. The difference between experiment and computation is compared and discussed in detail.

## 2 Experimental setup

### 2.1 Introduction to MNCP code

The Monte Carlo method is known as a random or statistical skills test method. MCNP is a universal package developed by Los Alamos National Laboratory based on the Monte Carlo method, and it has been used in calculating transportation problems of neutrons, photons, and electrons in three-dimensional complex geometries. MCNP can be used to calculate the properties of materials in three-dimensions (and even a special four-dimensional case). The program version slected here is MCNP5, which was released in 2003, compatible with existing versions 4C2/4C. It was significantly improved in the graphic display and convenience of installation as well as in the online documentation [18].

### 2.2 Modeling

#### 2.2.1 Calculation principle

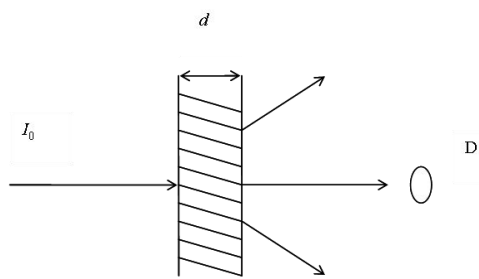

Fig. 1. Schematic diagram of narrow beam of rays

As shown in Fig. 1, a monoenergetic photon (Eγ) horizontally irradiates the surface of an absorber, which has a thickness d, and on the other side of the absorber there is a photon detector D. Photon transmission for the wide beam geometry obeys the following attenuation formula[19]:

$$I = BI_0 e^{-\mu_m \rho d} \quad (1)$$

where $I$ is the intensity after the photons pass through the material; $I_0$ is the intensity before the photons pass through the material; $B$ is the build-up factor for the wide beam geometry (determined by $I_0$, thickness $d$ and atomic number of the shield material); $\mu m$ is the mass attenuation coefficient of the shield, and ρ is the density of the shield.

### 2.2.2 Model specification

The experiment was carried out using the MCNP5 code. Spherical geometries were employed for the modeling of composite samples. The grid cards, surface cards and material cards of the MCNP5 program were used to build models and to set up the material type. The geometric model was divided as several lattice cells. The data cards include particle type, physical parameters of lattice cells, source and material description, result counting card, etc.

As shown in Fig. 2, the γ-ray source was located in the inner edge of the geometric model, and is a unidirectional point source. The distance from the source to the front of shielding material was set as 2 cm. The size of the shield material was set as 8 cm * 8 cm in cross section and 2 mm in thickness. During the simulation, the initial quantity of γ photons was set as $10^8$. The entire model is enveloped by a sphere filled with air inside and surrounded by vacuum outside. The sources were defined as point source, collimated beam, and monoenergetic source energies with uniform distribution of radioactive material (Cs-137, Am-241 and hypothetical γ-ray sources) upon them that emit γ-rays perpendicular to the front face of the shields and parallel to the direction of the sub-cylinders along the same axis (in the direction of the Z axis). The number of particles detected by the detector with and without the shielding material were calculated.

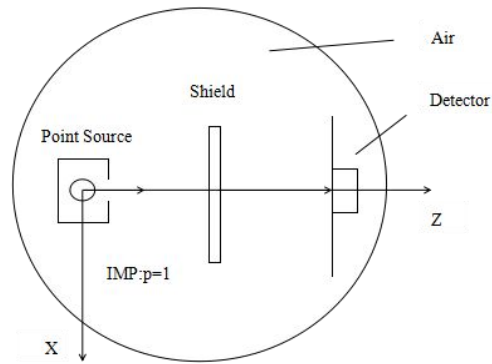

Fig. 2. X-Z sectional view of MCNP5 model

### 2.3 Experimental test of γ-ray shielding

Firstly, the functional filler i.e. PbWO$_4$ and Bi$_2$WO$_6$, and the substrate materials EPDM were applied to synthesize the composites. Among various polymers, EPDM is one of the ideal candidates, as it has low density (0.87 g/cm$^3$). PbWO$_4$ with mean particle diameter of about 4.5 μm was prepared by the precipitation method with surfactant MAA+PEG200. Bi$_2$WO$_6$ was synthesized by the hydrothermal process and the sample had an average particle size of 2.5 μm. The ball milled tungstate powders were homogeneously dispersed into the EPDM latex and then vulcanized .The loadings and the densities of the prepared composites are shown in Table 1.

Table 1. Densities of the composites prepared by PbWO$_4$ or Bi$_2$WO$_6$ with MCNP.

The thickness of each composites sheet is 2 mm.

| Contents/wt% | 0 | 10 | 20 | 30 | 40 | 50 |
|---|---|---|---|---|---|---|
| $\rho(PbWO_4)/g \cdot cm^{-3}$ | 0.984 | 1.173 | 1.256 | 1.414 | 1.530 | 1.636 |
| $\rho(Bi_2WO_6)/g \cdot cm^{-3}$ | 0.984 | 1.010 | 1.186 | 1.233 | 1.396 | 1.487 |

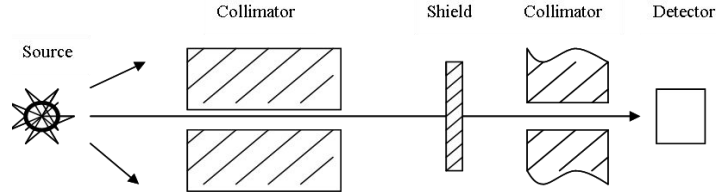

Fig. 3. The diagram of the experimental test of γ-ray shielding

The experimental tests of γ-ray shielding were carried out on a 3"X3" NaI + inspector 2000 with Am-241 and Cs-137 as the radioactive sources. As shown in Fig. 3, a slender hole was opened in the γ-ray-proof material such as lead bricks, thus the γ-ray was shaped into a beam. In this article the sectional area and thickness of each composite used in the experiment was 8cm*8cm and 2 mm, respectively. The distance between the detector and was 12 cm. The rubber sheet was inserted between the detector and γ-ray source and measured for 120 s. Each measurement was repeated 4 times and then the average was taken.

## 3 Results and discussion

### 3.1 Simulations of shielding performance

The mass attenuation coefficient ($\mu_m$) is a function related to the properties of materials and the energy of radiation. The results of simulations based on the selected model are showed in Table 2. The results show that the shielding performance of the two materials varied according to radiation source with different energy. It depends on the content proportion as well. The shielding performance of the composites for a γ-ray source with low energy (59.5 keV) increases significantly with the increase of the filling ratio of either $PbWO_4$ or $Bi_2WO_6$. However, the situation is different when the source energy is 661 keV, where there are no obvious changes in the shielding performance during the increase of the functional filler.

Table 2. The obtained mass attenuation coefficient ($\mu_m$) from simulations

| | Contents/wt% | 0% | 10% | 20% | 30% | 40% | 50% |
|---|---|---|---|---|---|---|---|
| 59.5KeV | $\mu_m(PbWO_4)$ | 0.3441 | 1.0482 | 1.1630 | 1.5153 | 1.7031 | 1.6790 |
| | $\mu_m(Bi_2WO_6)$ | 0.3441 | 0.5702 | 0.8059 | 1.2929 | 1.7792 | 2.1647 |
| 661KeV | $\mu_m(PbWO_4)$ | 0.1060 | 0.1785 | 0.1840 | 0.1673 | 0.1640 | 0.1561 |
| | $\mu_m(Bi_2WO_6)$ | 0.1060 | 0.2076 | 0.1905 | 0.1919 | 0.1811 | 0.1700 |

### 3.2 Comparison between experiment and simulation

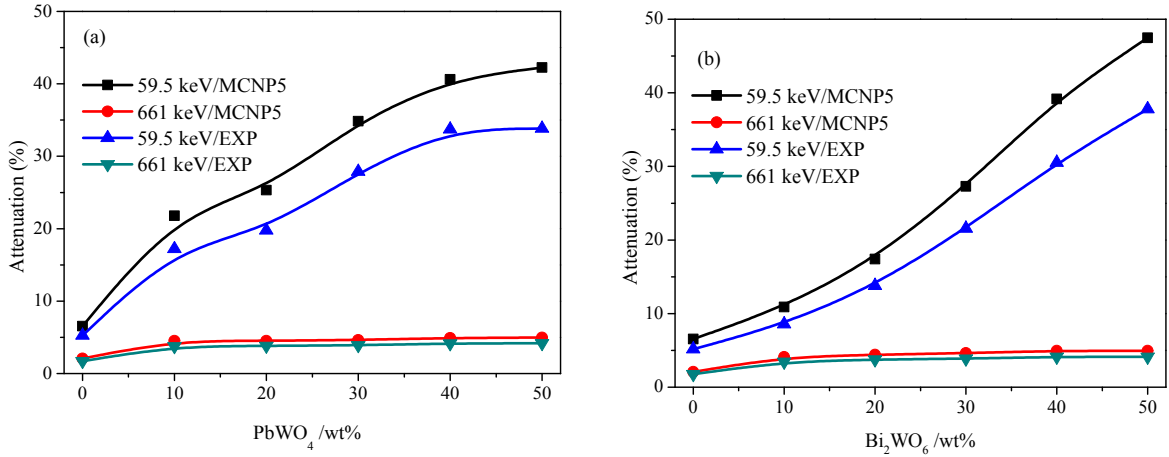

Fig. 4. Comparison of the shielding performances of PbWO$_4$ (a) and Bi$_2$WO$_6$ (b) by experiment (EXP) and simulation (MCNP5)

As shown in Fig. 4, PbWO$_4$/EPDM exhibits higher shielding capacity if a low proportion of functional filler is added (10-20 wt%). Taking the radiation energy at 59.5 keV as the example, the attenuation rates of PbWO$_4$/EPDM are about 20% when the loading of PbWO$_4$ is 10 wt%. The attenuation rates of Bi$_2$WO$_6$/EPDM decreased to around 10% even with the same mass loading of Bi$_2$WO$_6$. However, Bi$_2$WO$_6$/EPDM exhibits better shielding performance when the proportion of tungstate increased to 50 wt%. Further inspection of the curves shows that the trend of the shielding performances obtained from experiments and simulations are very consistent. There is a difference of about 10-20% between these two methods. The difference is probably due to the following reasons: 1) Apart from the photoelectric effect, Compton scattering was not considered in the simulation. 2) The geometry employed in the simulation was not perfectly identical to that in the experiment, which will result in certain differences. For instance, the effect of build-up factor was not excluded in simulation, but could play an important role for low energy radiation. 3) The emitted irradiation particles are not ideally uniform compared with those in the simulation process. 4) The likely material non-uniformity in the experiment can affect the results as well. In general, the difference is within the acceptable range. Tt reflects the different accuracies between the two methods and suggests that the MCNP code could be a potential candidate to calculate and predict the γ-ray shielding capacity of the functional materials.

### 3.3 Computational simulation of multi-layer γ-ray shielding material

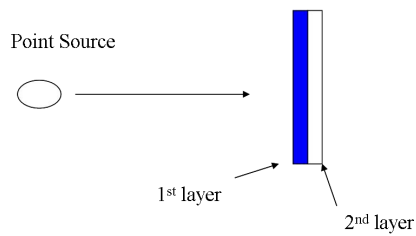

Fig. 5. The schematic of dual-layer shield model in simulation

The shielding parts used in nuclear reactors commonly consist of several layers of different materials [20]. It is necessary to study the influence of layer arrangement order and the properties of the multi-layer γ-ray shielding materials on the shielding capacity. Based on the above finding, we have further studied the multi-layer shielding systems by MCNP5 code, which is much more convenient than experimental testing. Thus, a double-layer model composed of two composites, i.e. PbWO$_4$/EPDM and Bi$_2$WO$_6$/EPDM, was designed. For each composite, the proportion of tungstate was 50 wt%. The thickness of all the materials systems was 2 mm.

Firstly, the shielding capacity of the single composite was investigated. It was found that $PbWO_4$/EPDM (labeled as Model A) shows higher shielding performance than that of $Bi_2WO_6$/EPDM (labeled as Model B), which is in accordance with the above results. Then the $PbWO_4$/EPDM (A) was located closer to the ray source with $Bi_2WO_6$/EPDM (B) on the other side, which was named Model A-B. Alternatively, when the position of these two materials was exchanged, the model was labeled as Model B-A (Figure 5). The thickness ratios of these two materials and the γ-ray used were varied. The selected thickness ratios of A to B were 3:1, 2:1, 1:1, 1:2, and 1:3. The energy of γ-rays used was varied from 60 keV to 660 keV.

The results are shown in Fig. 6. Both double-layer models exhibit higher shielding capacity than that of single $Bi_2WO_6$/EPDM or $Bi_2WO_6$/EPDM, even though all the systems have the same thickness of 2 mm. In addition, the B-A models (where $Bi_2WO_6$/EPDM is closer to the irradiation source) are more effective than the A-B models, no matter which kind of thickness ratios were used. The shielding performance can be dominant in γ-ray absorption [21]. The heavy metal elements like Pb, Bi, and W have different absorptive ranges, which will influence the γ-ray shielding performance. Here, a conclusion can be obtained that, for the $Bi_2WO_6$/EPDM to $PbWO_4$/EPDM double layer system, A as the back layer can be a better complementary absorber to B. The maximum attenuation rate was 58%, which was obtained when the thickness ratio of $PbWO_4$/EPDM (A) to $Bi_2WO_6$/EPDM (B) was 1 to 3.

In order to further study the influence of material arrangement on the γ-ray shielding, the K value based on the different irradiation sources was calculated. K equals the ratio of attenuation rate of the A-B model to that of the B-A model, which is observed to be smaller than 1 (Fig. 7). A bigger K value indicates that a smaller difference in the γ-ray shielding owing to the material arrangement. Taking the thickness ratio of A to B as 1 to 3 (square curve) as the example, when the photon energy was 60 keV, K was 0.886, and when the energy increased to 660 keV, K was 0.968, which means the smaller the photon energy is, the larger the difference is.

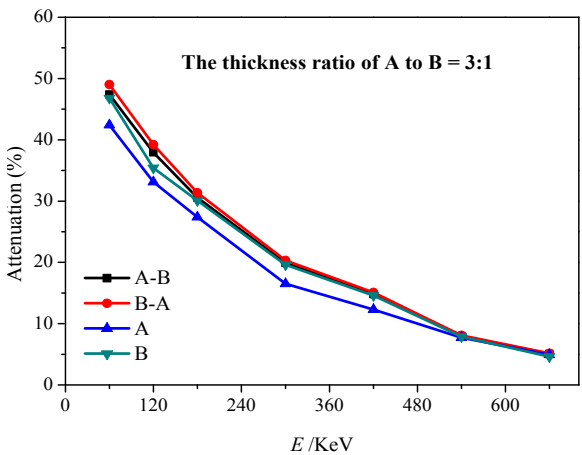
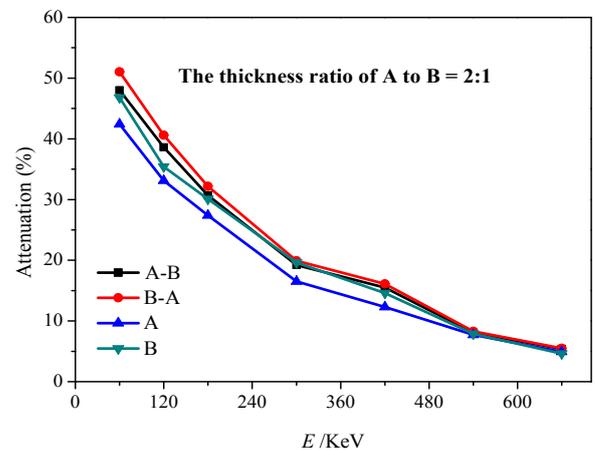

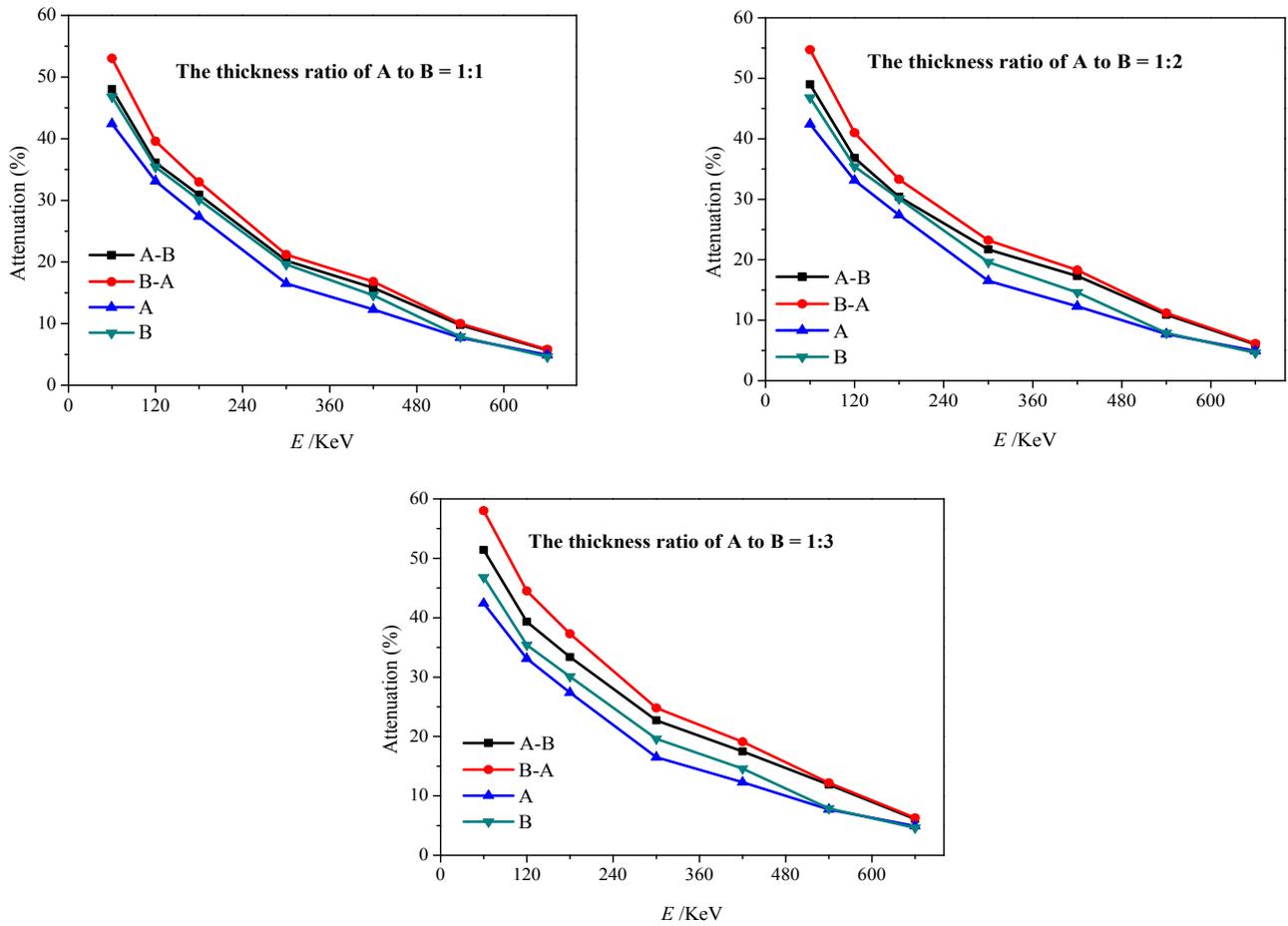

Fig. 6. The γ-ray shielding performances on the single and double-layer systems

It can also be observed that the K value was affected by the composition of the double-layer composites even if the photon energy was constant. Taking the photon energy of 60 keV as an example, when the thickness ratio of A to B was 3:1, K was 0.967 and when the thickness ratio of A to B was decreased to 1:3, the K value was 0.886, which means the smaller the ratio is, the larger the difference is. It has to be pointed out that K is always smaller than 1 in all the investigated cases.

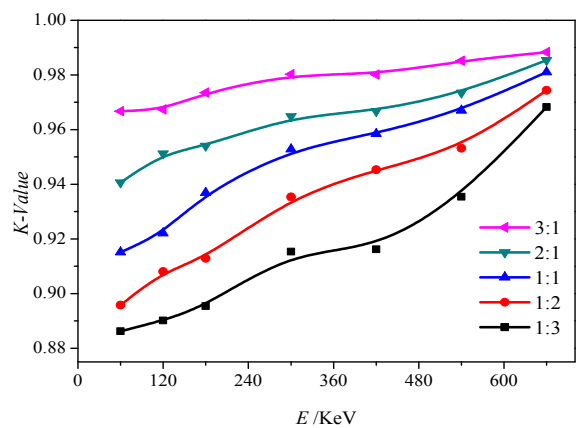

Fig. 7. The K-Value of the B-A model with different thickness ratios of A to B and different photon energies

## 4 Conclusion

In summary, $Bi_2WO_6$/EPDM and $PbWO_4$/EPDM composites with a thickness of 2 mm for γ-ray shielding were investigated by both experimental testing and computational simulation. The results from the different methods are strongly consistent. The results show that, with a 59.5 keV irradiation energy, $PbWO_4$/EPDM exhibited higher shielding capacity at a low loading (< 40 wt%). However, $Bi_2WO_6$/EPDM was better when the proportion of tungstate increased to 50 wt%. Further investigation of multi-layer systems by the MCNP5 code unveiled that double-layer models exhibited higher shielding capacity than that of either single $Bi_2WO_6$/EPDM or $Bi_2WO_6$/EPDM. In addition, double-layer materials with $Bi_2WO_6$/EPDM located closer to the irradiation source gave a better γ-ray shielding performance. The difference in the γ-ray shielding by the arrangements of the shielding materials can be influenced by the energy of the irradiation source, in which a higher energy leads to a small difference. The results suggest that the MCNP5 code could be a promising candidate to calculate and predict the γ-ray shielding capacity of functional materials.

## 5 Acknowledgement

*The authors are grateful for the financial support by the Research Fund of Southwest University of Science and Technology (15zx7159) and the Open Fund of State Key Laboratory Cultivation Base for Nonmetal Composites and Functional Materials, Sichuan Province (13zxfk07).*